\newif\ifsubmode
\submodefalse

\newif\ifprintfig
\printfigtrue

\newif\ifemulate
\emulatetrue

\ifsubmode
  \documentclass[12pt,preprint]{aastex}
  \received{}
  \accepted{}
  \journalid{}{}
  \articleid{}{}
\else
   \documentclass{emulateapj}
\fi
\usepackage{amssymb,amsmath,mathrsfs,hyperref,datetime}

\newcommand{\datavector}[1]{\boldsymbol{#1}}
\newcommand{\flux}{\datavector{F}}
\newcommand{\uncertainty}{\datavector{\sigma_F}}
\newcommand{\hyperpars}{\datavector{\alpha}}
\newcommand{\setofall}[1]{\left\{{#1}\right\}}
\newcommand{\allflux}{\setofall{\flux_i}}
\newcommand{\dd}{\mathrm{d}}

\def\urltilda{\kern -.15em\lower .7ex\hbox{\~{}}\kern .04em}

\pdfoutput=1

\begin{document}

\title{Star-galaxy classification in multi-band optical imaging}
\author{Ross~Fadely\altaffilmark{1},
        David~W.~Hogg\altaffilmark{2,3},
        \& Beth~Willman\altaffilmark{1}}
\altaffiltext{1}{Haverford College, Department of Physics and Astronomy, 370 Lancaster Ave., Haverford, PA, 19041}
\altaffiltext{2}{Center~for~Cosmology~and~Particle~Physics, Department~of~Physics, New~York~University, 4~Washington~Place, New~York, NY 10003, USA}
\altaffiltext{3}{Max-Planck-Institut f\"ur Astronomie, K\"onigstuhl 17, 69117 Heidelberg, Germany}

%
%
\begin{abstract}
  Ground-based optical surveys such as PanSTARRS, DES, and LSST, will
  produce large catalogs to limiting magnitudes of $r \gtrsim 24$.
  Star-galaxy separation poses a major challenge to such surveys
  because galaxies---even very compact galaxies---outnumber halo stars
  at these depths.  We investigate photometric classification
  techniques on stars and galaxies with intrinsic FWHM $<0.2$~arcsec.
  We consider unsupervised spectral energy distribution template
  fitting and supervised, data-driven Support Vector Machines (SVM).
  For template fitting, we use a Maximum Likelihood (ML) method and a
  new Hierarchical Bayesian (HB) method, which learns the prior
  distribution of template probabilities from the data.  SVM requires training 
  data to classify unknown sources; ML and HB don't.  We consider i.) a
  best-case scenario (SVM$_{best}$) where the training data is
  (unrealistically) a random sampling of the data in both
  signal-to-noise and demographics, and ii.) a more realistic scenario
  where training is done on higher signal-to-noise data
  (SVM$_{real}$) at brighter apparent magnitudes.  Testing with COSMOS
  $ugriz$ data we find that HB outperforms ML, delivering $\sim80\%$
  completeness, with purity of $\sim60-90\%$ for both stars and galaxies.  
  We find no algorithm delivers perfect performance, and 
  that studies of metal-poor main-sequence turnoff stars
  may be challenged by poor star-galaxy separation.  Using the Receiver Operating 
  Characteristic curve, we find a best-to-worst ranking of SVM$_{best}$, HB, ML, and 
  SVM$_{real}$.  We conclude, therefore, that a well trained SVM will outperform 
  template-fitting methods.  However, a normally trained SVM performs worse.  Thus, 
  Hierarchical Bayesian template fitting may prove to be the optimal classification method 
  in future surveys.  
\end{abstract}

%
%
\section{Introduction}

Until now, the primary way that stars and galaxies have been
classified in large sky surveys has been a morphological separation
\citep[e.g.,][]{kron80,yee91,vasconcellos11a,henrion11a} of point
sources (presumably stars) from resolved sources (presumably
galaxies).  At bright apparent magnitudes, relatively few galaxies
will contaminate a point source catalog and relatively few stars will
contaminate a resolved source catalog, making morphology a sufficient
metric for classification.  However, resolved stellar science in the
current and next generation of wide-field, ground-based surveys is
being challenged by the vast number of unresolved galaxies at faint
apparent magnitudes.  

\begin{figure}
\centering
 \includegraphics[clip=true, trim=0cm 0cm 0.0cm 0.cm,width=8cm]{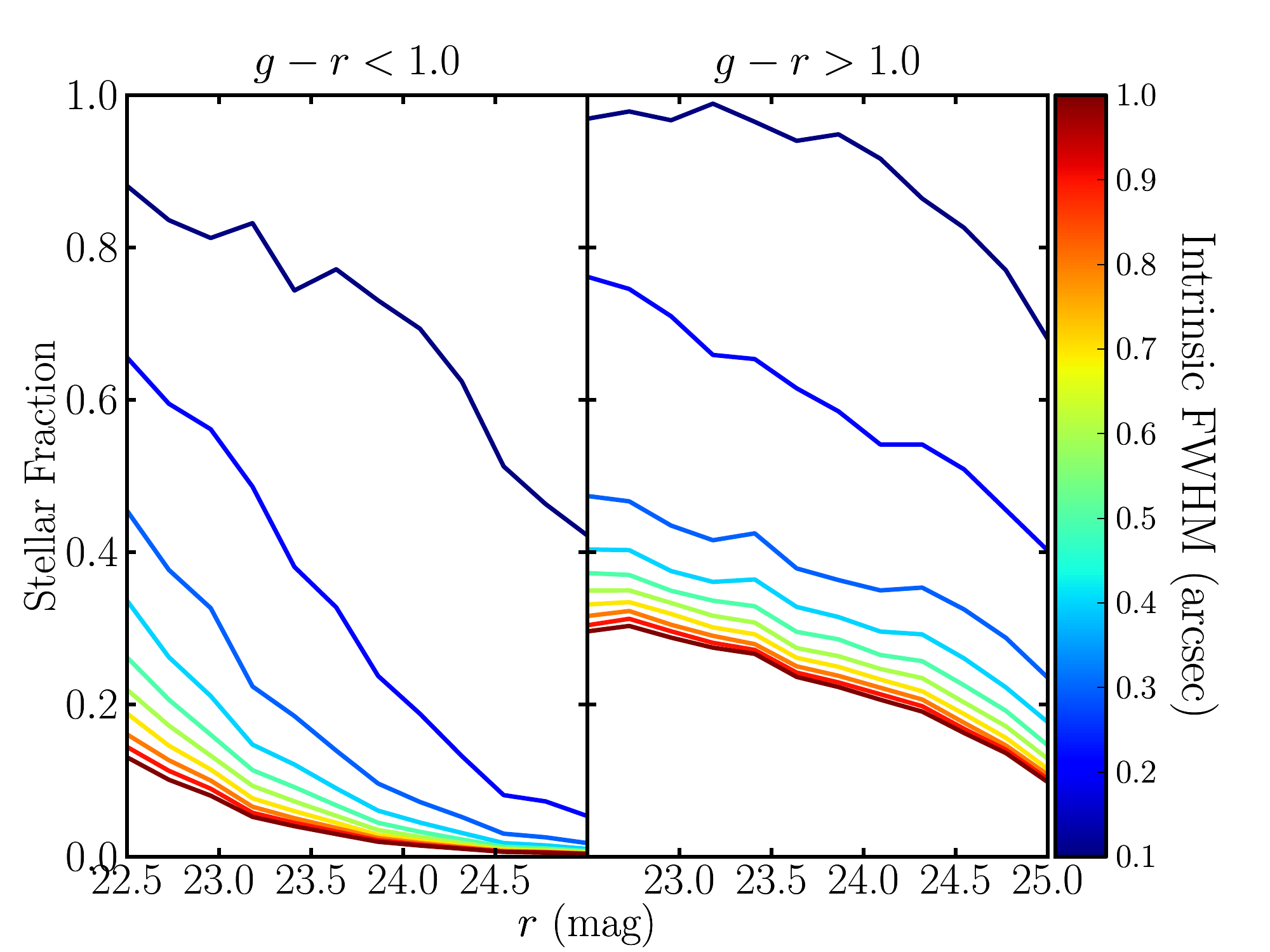}
\caption{The stellar fraction of COSMOS sources as a function of
  magnitude, for sources with $g-r<1$ (left) and $g-r>1$ (right).
  Only stars and galaxies were included in this figure; Only a few
  percent of the COSMOS point sources are AGN.  Colored curves
  indicate the upper limit in intrinsic full-width half-maximum
  (FWHM) allowed in the sample.  Even in an optimistic scenario where
  galaxies with FWHM $\gtrsim 0.2$~arcsec can be morphologically distinguished
  from stars, unresolved galaxies will far outnumber stars in point
  source catalogs at faint magnitudes.  This challenge is much greater
  for blue stars than for red stars.}
\label{fig:stellarfraction}
\end{figure}

To demonstrate this challenge for studies of field stars in the Milky Way (MW),
we compare the number of stars to the number of unresolved galaxies at
faint apparent magnitudes.  Figure~\ref{fig:stellarfraction} shows the
fraction of COSMOS sources that are classified as stars as a function of $r$
magnitude and angular size.  The COSMOS catalog \citep[($l,b$) $\sim$ (237,43)
degrees, ][]{capak07a,scoville07b,ilbert09} relies on 30-band photometry
plus HST/ACS morphology for source classification (see Section 4 for
details).  In Figure~\ref{fig:stellarfraction} we plot separately relatively 
bluer ($g-r < 1.0$) and redder ($g-r > 1.0$) sources because bluer stars are 
representative of the old, metal-poor main sequence turnoff (MSTO) stars 
generally used to trace the MW's halo while redder stars are representative of the
intrinsically fainter red dwarf stars generally used to trace the MW's
disk.  We will see that the effect of unresolved galaxies on these two
populations is different, both because of galaxy demographics and
because the number density of halo MSTO stars decreases at faint
magnitudes while the number density of disk red dwarf stars increases at
faint magnitudes.
 
In an optimistic scenario in which galaxies with FWHM $\gtrsim 0.2$~arcsec can
be morphologically resolved (the blue line in
Figure~\ref{fig:stellarfraction}, second from the top), unresolved
galaxies will still greatly outnumber field MW stars in a point source
catalog.  For studies of blue stars, field star counts are dominated
by unresolved galaxies by $r\sim23.5$ and are devastated by unresolved
galaxies at fainter magnitudes. The problem is far less severe for
studies of red stars, which may dominate point source counts for
$r\lesssim24.5$. Although morphological identification of galaxies with
FWHM as small as $0.2$~arcsec is better than possible for the Sloan Digital Sky
Survey (median seeing $\sim1.3$~arcsec), future surveys with
higher median image quality (for example, $0.7$~arcsec predicted for LSST) may
approach this limit.

Utilizing the fundamental differences between SEDs of stars and
galaxies can mitigate the contamination of unresolved galaxies in
point source catalogs.  In general, stellar SEDs are more sharply
peaked (close to blackbody) than galaxies, which exhibit fluxes more
broadly distributed across wavelength.  Traditionally, color-color
cuts have been used to eliminate galaxies from point source catalogs
\citep[e.g.,][]{gould92a,reitzel98a,daddi04a}.  Advantages of the
color-color approach include its simple implementation and its
flexibility to be tailored to the goals of individual studies.
Disadvantages of this approach can include its simplistic treatment of
measurement uncertainties and its limited use of information about
both populations expected demographics.

Probabilistic algorithms offer a more general and informative approach
to photometric classification.  The goal of probabilistic photometric
classification of an astronomical source is to use its observed fluxes
$\datavector{F}$ to compute the probability that the object is of a
given type.  For example, a star ($S$) galaxy ($G$) classification
algorithm produces the posterior probabilities $p(S|\datavector{F})$ and
$p(G|\datavector{F})$ and decides classification by comparing the
ratio of the probabilities
\begin{eqnarray}\displaystyle
\Omega = \frac{p(S|\datavector{F})}{p(G|\datavector{F})} 
\quad .
\label{eqn:oddsratio}
\end{eqnarray}
A natural classification threshold is an odds ratio,
$\Omega$, of 1, which may be modified to obtain more pure or more
complete samples.

Algorithmically there are a large number of approaches which produce
probabilistic classifications.  Generally, these fall into i)
physically based methods---those which have theoretical or empirical
models for what type of physical object a source is, or ii) data
driven methods---those which use real data with known classifications
to construct a model for new data.  Physically based Bayesian and
$\chi^2$ template fitting methods have been extensively used to infer
the properties of galaxies \citep[e.g.,][]{coil04a, ilbert09, xia09,
  walcher11a,hildebrandt10}. However, in those studies relatively
little attention has been paid to stars which contribute marginally to
overall source counts (although see \citealt{robin07}).  Several
groups have recently investigated data driven, support vector machine
based star--galaxy separation algorithms
\citep[e.g.,][]{saglia12,solarz12a,tsalmantza12a}.

In this paper, we describe, test, and compare two physically based
template fitting approaches to star--galaxy separation
(maximum-likelihood and hierarchical bayesian), and one data driven
(support vector machine) approach.  In Section 2, we present the
conceptual basis for each of the three methods.  In Section
\ref{sec:data}, we describe the COSMOS data set with which we test the
algorithms.  In Section \ref{sec:specifics}, we discuss the specific
details, choices, and assumptions made for each of our classification
methods. Finally, in Section \ref{sec:results} we show the performance
of the algorithms, and discuss the advantages and limitations related
to their use as classifiers.

%
%
\section{Probabilistic Photometric Classification Techniques}

\subsection{Template Fitting: Maximum Likelihood (ML)}
\label{ssec:MLmethod}

One common method for inferring a source's properties from
observed fluxes is template fitting.  This method requires a set of
spectral templates (empirical or theoretical) that span the possible
spectral energy distributions (SEDs) of observed sources.  These
template SEDs must each cover the full wavelength range spanned by the
photometric filters used to measure the fluxes to be fit.  The
relative template flux in each filter (for example $ugriz$) for each
SED is computed by convolving each SED with each filter response
curve.  Once these relative flux values are computed for
each SED template, the template model is fully specified except for a
normalization constant $C$.  For a given observed source $i$, the
value of $C_i$ is proportional to the total luminosity of the source
divided by the luminosity distance squared.  This value of $C_i$ is
unknown but can be `fit' to the data.

The maximum likelihood (ML) value of $C_i$ for each template that best
fits a source's observed fluxes, $\datavector{F}$, is that which
returns the lowest $\chi^2$. After assessing the ML values of $C_i$
for all the templates, classification is straightforward---one need
only to compare the lowest star $\chi^2$ to the lowest galaxy
$\chi^2$.  In other words, $\chi^2_S-\chi^2_G=\ln(\Omega)$ is the
classification criteria (see Equation \ref{eqn:oddsratio}).

\subsection{Template Fitting: Hierarchical Bayesian (HB)}
\label{ssec:HBmethod}

Hierarchical Bayesian (HB) algorithms provide another template
fitting-based approach to photometric classification.  Unlike ML
approaches, Bayesian approaches offer the opportunity to utilize
information about how likely a source is to be each kind of star or
galaxy; the different templates are not treated as equal \emph{a
priori}.  With a hierarchical Bayesian algorithm, individual source
prior probabilities do not need to be set in advance of the
full-sample classification process; the entire sample of sources can
inform the prior probabilities for each individual source.

Consider the scenario where a $G$ model fits data $\datavector{F}_i$
only \textit{slightly} better than the best $S$ model, while all other
$G$ models give poor fits and all other $S$ models give nearly as
likely fits.  In this case, ignoring all other $S$ models besides the
best is the wrong thing to do, since the data are stating that $S$
models are \textit{generally} more favored.  Capturing this kind of
information is one primary aim of most Bayesian algorithms.

To capture this information, we \textit{marginalize} over all possible
star and galaxy templates to compute the total probability that a
source belongs to a certain classification ($S$ or $G$).  For a
template fitting-based Bayesian algorithm, this marginalization
consists of summing up the likelihood of each $S$ template given
$\datavector{F}_i$, as well as the likelihood of each $G$ template
(across redshift).  Note that the likelihood of each template is itself
calculated as a marginalized likelihood.  For each template fit, we
compute the total likelihood of the fit by marginalizing over the
uncertainty in fitting coefficient $C_i$.  This marginalization is the
total probability of a Gaussian distribution with variance
$\sigma_{C_i}^2$---a value which is returned using least squares
fitting techniques \citep[e.g.,][]{hogg10a}.

By Bayes' theorem, marginalization requires we specify the prior
probability that any object might have a given SED template (at a
given redshift).  The prior probability distributions might be chosen
to be uninformative (for example, flat), informed by knowledge from
outside studies, or informed by the data on all the other objects.
The latter approach, referred to as a hierarchical model, is widely
used in statistical data analysis \citep[e.g.,][]{gelman03} and is
beginning to be used in astronomy \citep{mandel09,hogg10b,mandel11,
shu12}.  The benefits of hierarchical
approaches are many---because every inference is informed by every
datum in the data set, they generally show improved probabilistic
performance over simpler approaches, while requiring no additional
knowledge outside the observed data and the template SEDs.
Functionally, hierarchical approaches consist of parameterizing the
prior probability distributions (for example, with the mean and
variance of a Normal distribution), and varying these parameters
(known as ``hyperparameters'') to determine the probability
of \emph{all} the data under \emph{all} the models.

For our work, we optimize the hyperparameters of the SED template
prior distributions to return the maximum marginalized likelihood of
all the data.  This procedure will enable us to simultaneously infer
the star--galaxy probability of each source while determining the
hyperparameters that maximize the likelihood of the observed dataset.
A brief description of the functional form of these priors is given
below in Section~\ref{ssec:HBspecifics}.  Although we focus on the
star--galaxy probabilities in this paper, the optimized hyperparameters
themselves yield a measurement of the detailed demographics of a
dataset. 

\subsection{Support Vector Machine (SVM)}
\label{ssec:SVMmethod}

A support vector machine (SVM) is a type of machine learning algorithm
particularly well suited to the problem of classification.  SVM
algorithms are frequently used in non-astronomical problems, and are
considered a gold standard against which to compare any new
classification method.  SVM algorithms are ``supervised'', meaning
they train on a catalog of objects with known classifications to learn
the high dimensional boundary that best separates two or more classes
of objects.  For classification problems which do not separate perfectly, 
SVMs account for misclassification errors by looking at the degree of 
misclassification, weighted by a user specified error penalty parameter.  
In general the optimal boundary need not be restricted to a linear hyperplane, 
but is allowed 
to be non-linear and so can require a very large number of parameters to 
specify the boundary.  In order for non-linear SVM classification to be 
computationally feasible, a kernel function is used to map the problem to a 
lower dimensional feature space \citep{boser92}.

For the case of star--galaxy separation based on broad
band photometry, the SVM algorithm learns the boundary which best
separates the observed colors and apparent magnitudes\footnote{We use
  apparent $r$ magnitude here.} of stars and galaxies.  For more
details on the SVM technique, please see \citet{muller01}.

Successful implementation of a SVM algorithm requires a training dataset
that is a sufficient analog to the dataset to be classified.  A SVM has
recently been applied to source classification in the Pan-STARRS 1
photometric pipeline \citep{saglia12}, with promising initial
results. However, these results were obtained based on analysis of 
bright, high signal-to-noise data ($r\lesssim18$), using training data 
which is a subset of the data itself.

To investigate the impact of training set quality and demographics on
the problem of star--galaxy separation, we will consider the utility
and performance of SVM algorithms in a new classification regime,
where the data is of lower signal to noise (described in Section
\ref{sec:data}), and the number of unresolved galaxies is comparable
to or larger than the number of stars.

%
%

\section{Test Data}
\label{sec:data}

To investigate the advantages and disadvantages of star--galaxy
classification techniques, we need a test catalog which has a large
number of sources, is well understood and calibrated, and for
which spectroscopy or multi-wavelength observations reveal the true
source classifications.  In addition, we want these data to be
magnitude limited as faint as $r\ge24$ in order to understand the
problem of classification in current and upcoming surveys like
Pan-STARRS 1, DES, and LSST.  The COSMOS catalog satisfies these
requirements.   

The \href{http://cosmos.astro.caltech.edu/}{COSMOS} survey 
\citep{scoville07a} covers $\sim$ 2 square degrees on
the sky using 30 band photometry, and is magnitude limited down to $r
\sim 25$.  Broadband $ugrizJK$ photometry exists down to limiting
magnitudes which complement the $r$ limiting magnitude, and {\it
  Spitzer} IRAC coverage exist for sources as faint as $K\lesssim24$ 
  \citep{capak07,sanders07,taniguchi07}.
In addition, {\it GALEX} and {\it XMM} coverage are of sufficient
depth to pick out relatively bright star-forming galaxies and AGN 
\citep{hasinger07,zamojski07}.
The spectral coverage beyond the optical, particularly the
near-infrared, can be a powerful discriminator between star and
galaxy classification.  For instance, \citet{ilbert09} show the
$r-m_{\rm 3.6\mu\,m}$ vs. $r-i$ colors cleanly separate star and
galaxy loci, since stars have systematically lower $r-m_{\rm
  3.6\mu\,m}$ colors.  In addition to 30 band photometry, the COSMOS
field has \textit{HST/ACS} $i-$band coverage, down to a limiting
magnitude of $i\sim28$ \citep{koekemoer07,scoville07b}.  Diffraction
limited \textit{HST} imaging allows the morphological discrimination
of point-like and extended sources, further strengthening the fidelity
of the COSMOS star--galaxy classification.

We follow the COSMOS team's star--galaxy classification criteria 
in order to determine the `true' classification for the purpose of 
testing our methods.  These consist of a $\chi^2$ classification 
from fitting star and galaxy templates to the 30 band photometry, 
and a morphological classification using the 
{\tt ACS\_MU\_CLASS} statistic derived by the analysis of the 
\textit{HST} photometry by \citet{scarlata07}.   We label COSMOS
sources as stars if {\tt ACS\_MU\_CLASS} says the source is pointlike, 
and the `star' $\chi^2$ is lower than that for `AGN/QSO'.  For galaxies, 
we require the source to have a non-pointlike {\tt ACS\_MU\_CLASS}.  
This classification assumes that all galaxies in the \textit{HST} images 
are resolved.  We view this as an excellent approximation of the truth 
-- COSMOS \textit{ACS} images are very deep ($i\sim28$), and can thus detect 
the faint extended features of nearly unresolved galaxies.  We have 
qualitatively confirmed this by examining the distribution of galaxy 
FWHM, and find the distribution to be smoothly decreasing down to 
the smallest FWHM in the data.  We estimate the number of galaxies 
labeled as stars to be below the few-percent level.  For the labeling, we 
use an updated version of the publicly available photometric catalog, 
provided by P. Capak (private communication).  While present in the catalog, 
we do not use any photometric redshift information in determining the 
classification of COSMOS sources.

\begin{figure}
\centering
 \includegraphics[clip=true, trim=0cm 0cm 0.0cm 0.cm,width=8cm]{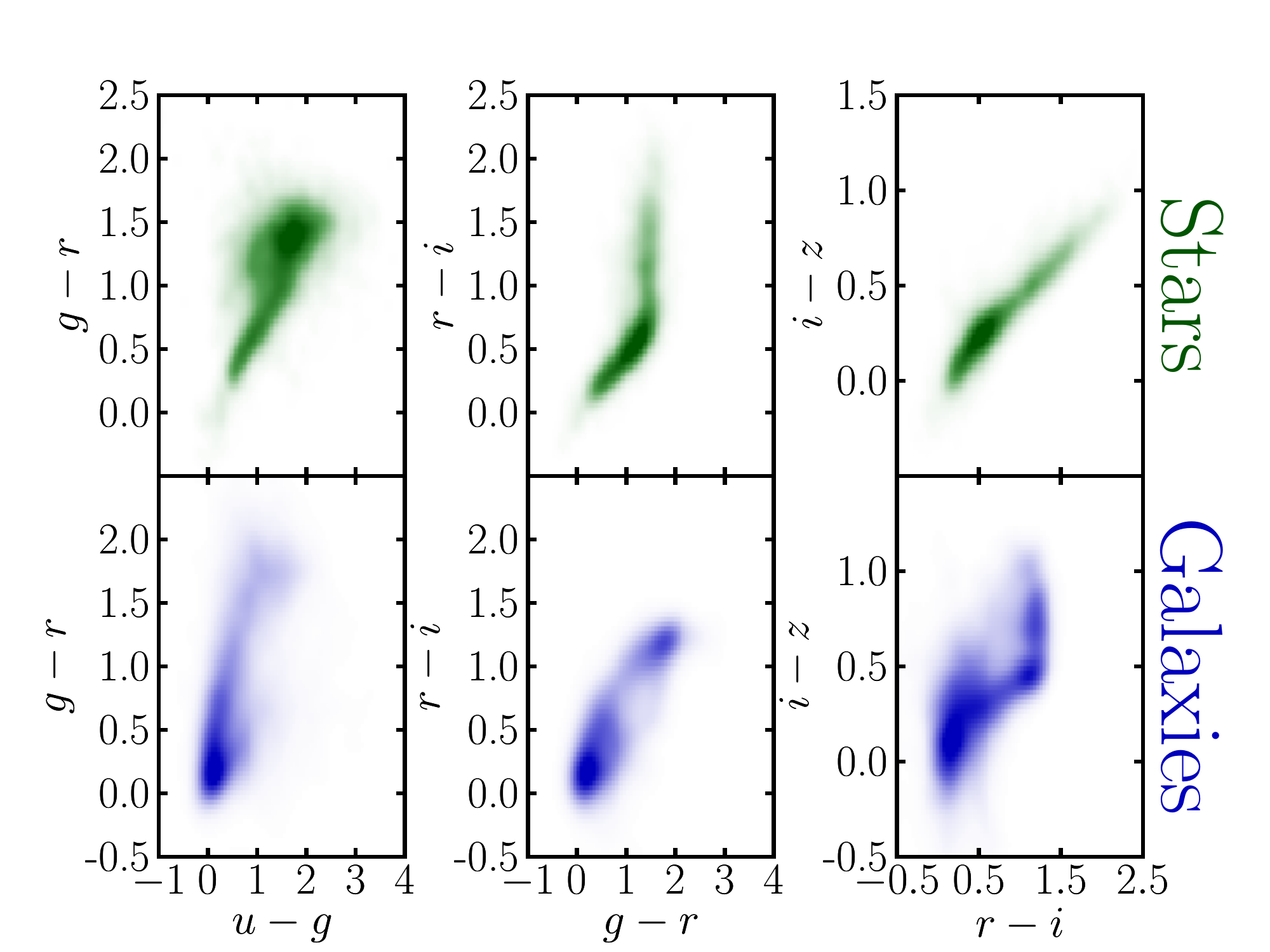}
\caption{The color--color space distribution of point sources (FWHM $<
  0.2$~arcsec) in the COSMOS catalog.  It is clear that stars in the sample
  follow a tight locus in all slices of color--color space, while
  galaxies are more generally distributed.  Even so, comparison by eye
  reveals significant overlap between stars and galaxies, particularly
  for bluer sources.}
\label{fig:color-color-data}
\end{figure}

Throughout this paper, we restrict our analysis to sources likely to
be unresolved in ground based data (FWHM$_{HST/ACS}$ $< 0.2$~arcsec).  We
do so since commonly used morphological classification criteria will
easily distinguish quite extended sources, accounting for a majority
of galaxies to depths of $r\sim24-25$.  However, galaxies with angular
sizes $< 0.2$~arcsec are unlikely to be resolved in surveys with seeing
$\gtrsim0.7$~arcsec, and so are an appropriate test bed for the type of sources
which will rely the most on photometric star--galaxy separation.  In
total, our sample consists of 7139 stars and 9167 galaxies with apparent 
magnitudes $22.5<r<25$, and is plotted in {\it ugriz} color--color space in
Figure \ref{fig:color-color-data}.  Over this magnitude range, the median 
signal-to-noise in the $r$ band ranges from $\sim50$ at $r=22.5$ to 
$\sim15$ at $r=25$, with lower corresponding ranges of 10 to 7 in the $u$.  
Of all 18606 sources with FWHM$<0.2$~arcsec in the COSMOS catalog, we 
identified 2300 AGN, which we discard from our current analysis. 

%
%

\section{Implementation of Three Star--Galaxy Classifiers}
\label{sec:specifics}

In this Section, we describe our implementation of ML template
fitting, HB template fitting, and a SVM on the $ugriz$ photometry of
COSMOS sources for purposes of star--galaxy classification.

\subsection{ML Template Fitting}
\label{ssec:MLspecifics}

Template based star--galaxy classification relies on the use of
spectral energy distribution templates which (as well as possible)
span the space of colors for both stars and galaxies.  For our stellar
model library, we first adopt the \citet{pickles98} set of empirically
derived SEDs, which span O to M type stars for both main sequence,
giant, and supergiant stars.  The vast majority of the SEDs in the
Pickles library have solar abundances, so we supplement the library
with theoretical SEDs from Castelli-Kurucz (CK) \citep{castelli04a}.
We use CK models with abundances ranging from $-2.5
\le\,$[Fe/H]$\,\le0.0$, surface gravities ranging from
$3.0\le\log(g)\le0.0$, and effective temperatures from $3500 \le
T_{eff} \le 10000\,$K.  We include binary star templates by combining
like-metallicity templates using flux calibrated CK models.  Finally, we include SDSS
M9 through L0 dwarf templates provided by J. J. Bochanski (private 
communication).  These
templates have been extended from the templates of \citet{bochanski07} into 
the near infrared, but lack data for wavelengths shorter than $4000\,$\AA.  We
extend these templates down to the $3000\,$\AA\, using a main sequence
CK model with $T_{eff}=3500$K.  Details of this extension are likely
to be unimportant, since the flux of such stars between
$3000-4000\,$\AA\, is negligible.  Our final combined library of stellar
templates includes 131 from the Pickles library, 256 from the CK library,
11 from \citet{bochanski07}, and 1319 binary templates constructed from
the CK library, for a total of 1717 stellar templates.

We select for our galaxy templates those used by the COSMOS team,
described in \citep{ilbert09}, provided publicly through the
\href{http://www.cfht.hawaii.edu/\%7Earnouts/LEPHARE/lephare.html}
{\texttt{Le Phare}} photometric redshift package\footnote{
\href{http://www.cfht.hawaii.edu/\%7Earnouts/LEPHARE/lephare.html}
{\tt http://www.cfht.hawaii.edu/\%7Earnouts/LEPHARE/lephare.html}}
\citep{arnouts99,ilbert06}.  These templates consist of galaxy SEDs
from \citet{polletta07}, encompassing 7 elliptical and 12 spiral
(S0-Sd) SEDs.  Additionally, 12 representative starburst SEDs are
included, which were added by \citet{ilbert09} to provide a more
extensive range of blue colors.  Templates from \citet{polletta07}
include effects of dust extinction, since they were selected to fit
spectral sources in the VIMOS VLT Deep Survey \citep{lefevre05}.  We
do not consider any additional dust extinction beyond these fiducial
templates.  In order to model our galaxies across cosmic time, we
redshift these templates on a discrete linear grid of redshifts,
ranging from 0 to 4 in steps of 0.08.  Simple tests using the ML
procedure indicate small changes to the step size of our grid are
unimportant. 

For all of the above templates, model fluxes were constructed by 
integrating the SED flux density values with the throughput response 
curves for each filter.  These consist of a $u^\ast$ response curve for 
the observations taken by the Canada-France-Hawaii Telescope, and 
$g^+$, $r^+$, $i^+$, $z^+$ response curves for data 
collected by the Subaru telescope.  We obtained the same response 
curves used by \citet{ilbert09} through 
\href{http://www.cfht.hawaii.edu/\%7Earnouts/LEPHARE/lephare.html}
{\texttt{Le Phare}}\footnotemark[5].  To check for any mismatch between 
the data, calibrations, and/or response curves, we verified that model 
colors generated from the SEDs overlap well with the star and galaxy loci.

\subsection{HB Template Fitting}
\label{ssec:HBspecifics}

While the HB template fitting technique builds on the foundation
described in Section~\ref{ssec:MLspecifics}, the details of
star--galaxy inference require significantly more mathematical
formalism to thoroughly describe.  We present the details of this
formalism and a detailed, step-by-step description of our HB
inferential procedure in Appendix A.  Open-source \texttt{C} code is
available at {\footnotesize
  \texttt{\url{http://github.com/rossfadely/star-galaxy-classification}}}.
In this section, we qualitatively describe features specific to our HB
algorithm.  We emphasize that hierarchical Bayesian algorithms are
unsupervised: we use no training set and do not set priors in advance
of running the algorithms.  As described in Section \ref{ssec:HBmethod}, 
the priors for the templates are inferred from the data itself.

Our HB template fitting method draws from the same set of SED
templates described above in Section \ref{ssec:MLspecifics}.  However,
to speed up the algorithm, we used only 250 of the 1313 star
templates, spanning a range of physical and color-color properties.  In 
practice, we find the individual choice of these templates to be 
unimportant (since many are very similar) so long as the templates span 
the colors of stars, with a sampling close to or better than the typical color 
uncertainties of the data.  We believe similar arguments to be true for 
galaxies, but have not explored such issues since we currently use only 
31 galaxy templates.

The primary choice we must make for our HB approach is the functional
form(s) of the prior probability distributions in the model.  Since
our templates are discrete both in SED shape and physical properties,
we parameterize the prior probability of each template to be a single
valued weight, within the range 0 to 1, such that the weights sum to
1 (see, for example, \ref{eqn:tempmarg} and \ref{eqn:tempconstraint}).  
These weights themselves become hyperparameters in our
optimization.  We thus have 281 hyperparameters corresponding to
template priors since we use 250 star and 31 galaxy templates.  The
overall prior probability that any given object is $S$ or $G$ is also
parameterized as two weights that sum to one (\ref{eqn:fullprob} and 
\ref{eqn:SGpriorconstraint} in Appendix), which we optimize.

For the galaxy models, we must choose a form for our redshift priors.  Ideally,
these should be parameterized as weights for each discrete redshift,
repeated as a separate set for each galaxy template.  Unfortunately,
this would not only add $51\times31$ more hyperparameters to optimize,
but also significantly slows down likelihood computations.  Instead,
we adopt a flat prior distribution across redshifts.  While not ideal,
such a prior eases comparison with ML classification results, and
eliminates the need to specify an informative prior which correctly
describes the data.  Tests of flat versus fixed-form prior
distributions indicate that the classification results presented in
Section \ref{sec:results} do not vary substantially between the two
choices.  

Finally, for each template fit we marginalize over the (Gaussian) uncertainty 
in the fit amplitude, for which we must specify a prior distribution 
(\ref{eqn:fitmarg} and \ref{eqn:fitconstraint} in Appendix).  We adopt a log-normal 
prior for the fit amplitudes, which we set by taking the mean and variance 
of the log-amplitudes from fits of all the data for a given template.  This approach makes 
the priors essentially uninformative, since the variance for all the data is 
large with respect to the variance for data which is well fit by the template.
Like redshift priors, these too could be treated as hyperparameters but come 
at the cost of much slower likelihood computations.  

In summary, we fix redshift and fit-amplitude priors and vary the prior weights 
of the template and $(S,G)$ probabilities.  Thus, we optimize 283 prior
(hyper)parameters to values which yield the maximum likelihood of the entire 
dataset.

\subsection{SVM Models}

We use the
\href{http://www.csie.ntu.edu.tw/\%7Ecjlin/libsvm/}{\texttt{LIBSVM}}
\footnote{\href{http://www.csie.ntu.edu.tw/\%7Ecjlin/libsvm/} {\tt
    http://www.csie.ntu.edu.tw/\%7Ecjlin/libsvm/}}| set of routines,
described in \citet{chang11a}.  The provided routines are quick and
easy to implement, and only require the user to specify a training set
of data, a set of data to be classified (a.k.a., test data), and the form 
and parameter values of the kernel function used.  

We employ a Gaussian radial basis function for the SVM kernel, 
for which we must specify a scaling factor $\gamma$.  Together with 
the error penalty parameter ($C_{\rm SVM}$) we have two 
nuisance parameters whose optimal values we need 
find.  We do this by using a Nelder-Mead simplex optimization algorithm to find the 
parameter values which provide the highest number of correct 
classifications in the test data.  In detail, the optimal values for 
$\gamma,C_{\rm SVM}$ will be different for each combination of 
training and test data.

To select the training data, we consider two scenarios.  First is a
`best case' situation (SVM$_{best}$), where a well-characterized
training set exists which is a fair sampling of the test data, with both 
the same object demographics and same signal-to-noise ($S/N$) as the data to be 
classified.  To emulate this scenario, we select
the training set as a random sample of the COSMOS catalog.  Second, we
consider a more realistic case where the available training set is
only sampling the demographics of the high $S/N$
portion of the catalog to be classified (SVM$_{real}$).  In this
case, the demographics of objects in the training set may not match
the demographics of the majority of objects in the set to be
classified.

We consider SVM$_{best}$ an optimistic scenario---obtaining a large
spectroscopic or multi-wavelength sample of training data, down to the
limiting magnitude of a given survey, is very costly in terms of
telescope time.  The other extreme, SVM$_{real}$, is a bit more
realistic---for a given survey, classifications are typically easily
obtained only at the high $S/N$ end of the data.  In both cases, we
consider a training sample size which is a fifth of the total catalog
size.

Finally, to implement the SVM classification routine we need to scale
both the training data and test data.  That is, for the colors and apparent magnitude used, we must
scale the range of each to lie between $-1$ and $1$.  We map both
training and test data to the interval $[-1,1]$ using the full range
of values in the test data.  This is important in the case of
SVM$_{real}$, since the training data may not span the full range of
values for the test data.  We find that scaling can have a significant
effect for the SVM$_{real}$ model.  For example, poor classification
performance is obtained if the SVM$_{real}$ training data is scaled to
itself rather than to the test data.

%
%

\section{Results and Discussion}
\label{sec:results}

\begin{figure}
\centering
 \includegraphics[clip=true, trim=0cm 0cm 0.0cm 0.cm,width=8cm]{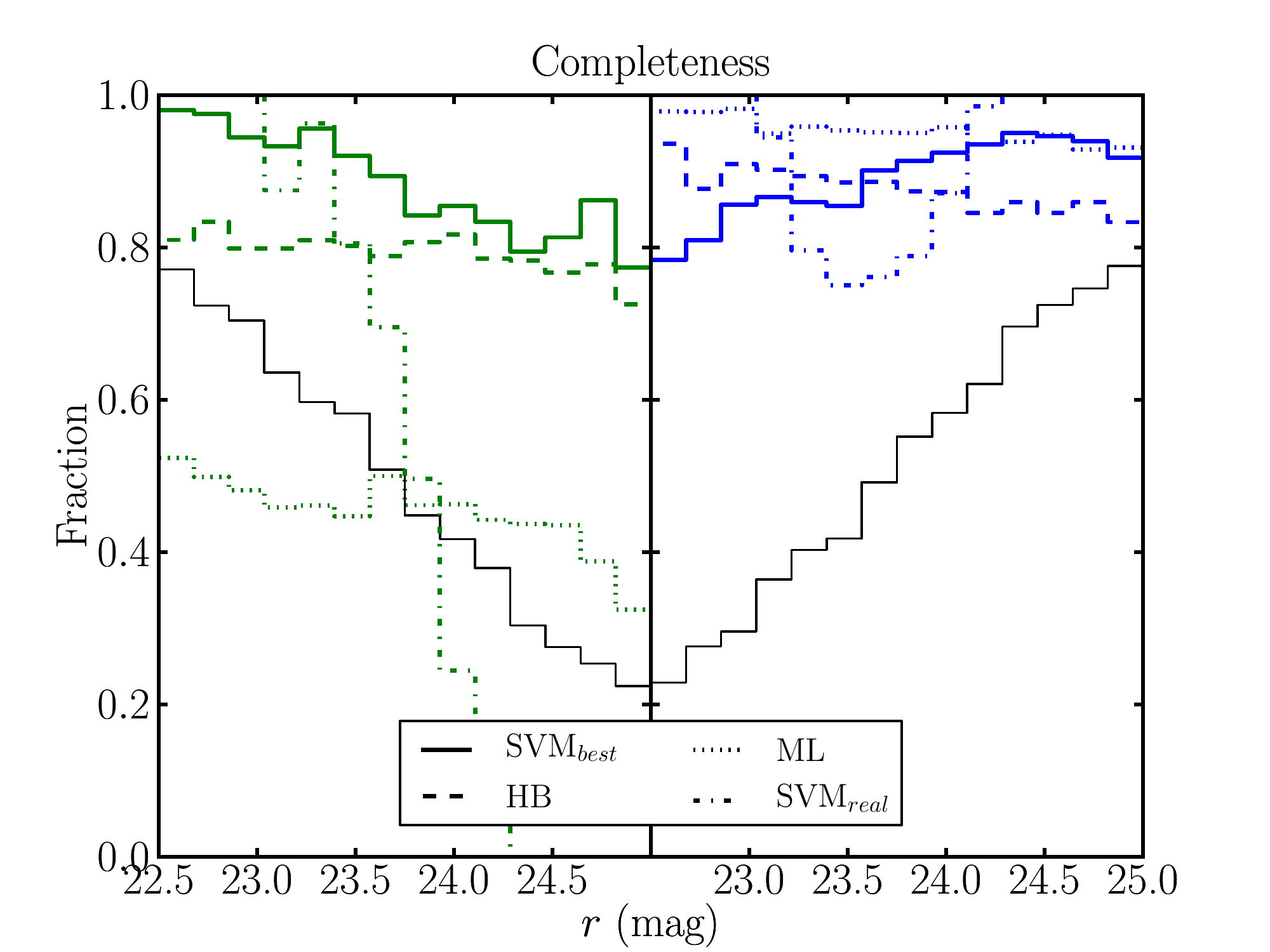}
\caption{The completeness as a function of magnitude produced by the
  indicated classification approaches.  Results for stars are on the
  left in green, while those for galaxies are shown on the right in
  blue.  The thin, solid black line indicates the sample fraction for
  a given object type.  For galaxies, the various methods return
  similar completeness values, while the discrepancy is much larger in
  the case of stars.}
\label{fig:completeness}
\end{figure}

\begin{figure}
\centering
 \includegraphics[clip=true, trim=0cm 0cm 0.0cm 0.cm,width=8cm]{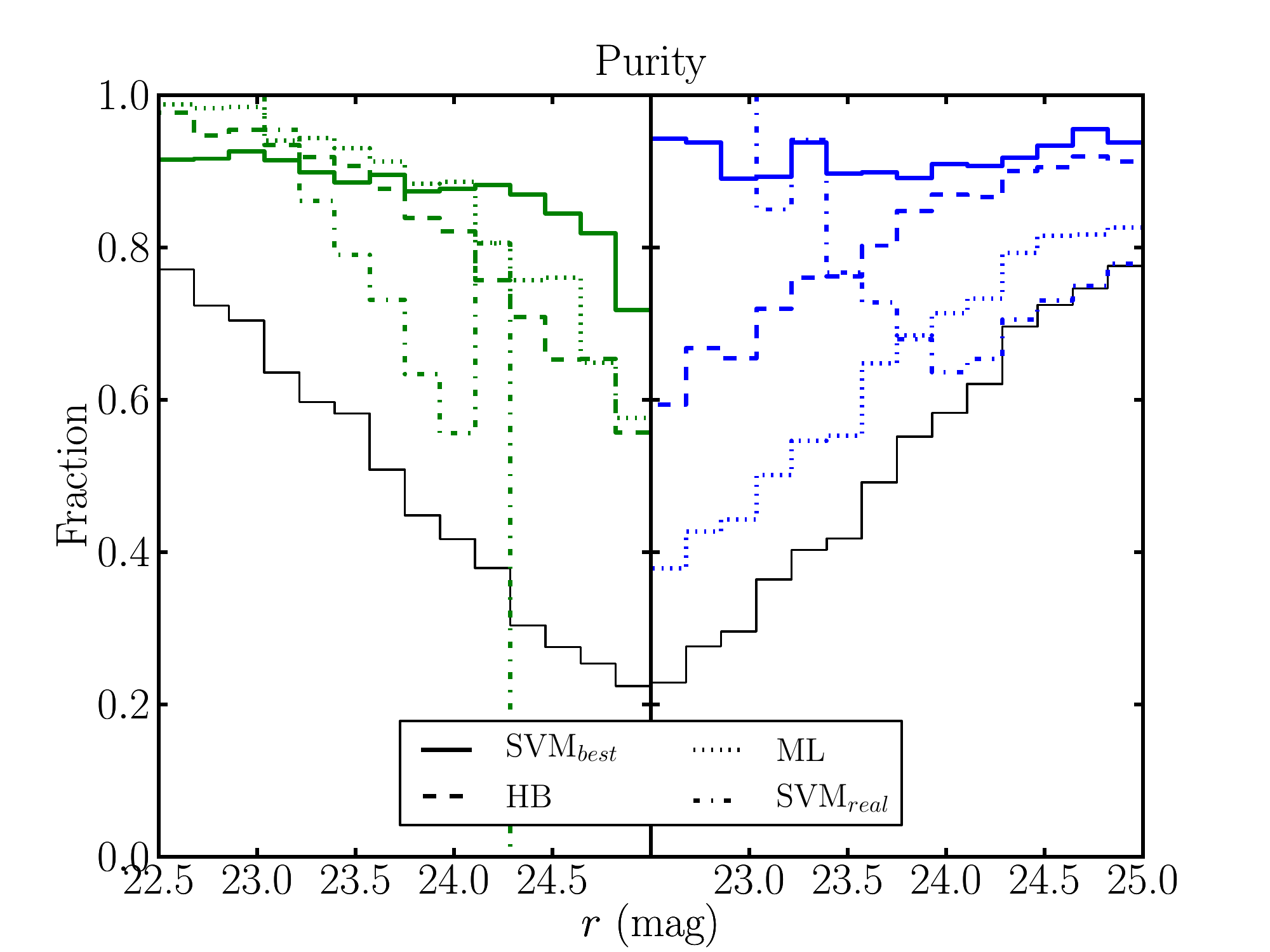}
\caption{Similar to Figure \ref{fig:completeness} but showing purity
  of classified samples, instead of completeness. Results for stars
  are on the left in green, while those for galaxies are shown on the
  right in blue.  Here, SVM algorithms generally outperform all others,
  if given a very good set of training data (${\rm SVM}_{best}$).  For
  stars, our HB algorithm delivers somewhat similar purity to the ${\rm
    SVM}_{best}$ scenario.  For galaxies, however, HB underperforms ${\rm
    SVM}_{best}$ as the stellar fraction of the sample decreases.}
\label{fig:purity}
\end{figure}

We report the classification performance of Maximum Likelihood (ML)
and Hierarchical Bayesian (HB) template fitting, as well as a
thoroughly tested Support Vector Machine (SVM) on our COSMOS based
test data.  There are many different measures which can be used to
assess the performance of each algorithm.  First, we consider the
completeness\footnote{Defined as the fraction of sources of true type
  $X$, correctly classified as $X$.} and purity\footnote{Defined as
  the number of sources of true type $X$, correctly classified as $X$,
  divided by the total number of sources classified as $X$.} of
classified samples, evaluated at $\ln(\Omega)\footnote{Defined in
  Equation 1}=0$.  Figures \ref{fig:completeness} and
\ref{fig:purity}, display the completeness and purity, respectively,
as a function of magnitude.  Examining Figure \ref{fig:completeness},
all methods seem to be fairly competitive for galaxy classification,
returning $80-90\%$ completeness across all magnitudes.  SVM$_{best}$
and ML yield the most consistently robust completeness for galaxies.
In the case of stars, however, it is clear only our HB template
fitting and SVM$_{best}$ deliver acceptable completeness---at $r>24$
the completeness of ML template fitting falls to 50\% or below, and
the completeness for SVM$_{real}$ goes to zero.  The mismatch in
source demographics between the realistic training set and the faint
COSMOS sources severely undermines the efficacy of SVM$_{real}$.

In terms of purity (Figure \ref{fig:purity}), SVM$_{best}$ outperforms
all other approaches.  For stars, HB yields similar performance to
SVM$_{best}$, but all approaches underperform SVM$_{best}$ in terms of
galaxy purity.  When taken in concert with the results of Figure
\ref{fig:completeness}, we see that HB delivers similar or better
performance than ML in all cases, even with the relatively simple HB
approach presented here.  For stars, ML and HB yield similar sample
purity, but HB does so with a much higher completeness ($\sim 80\%$
vs. $\sim 50\%$).  For galaxies, HB yields a consistently higher
sample purity by $\sim 10-15\%$ but a consistently lower sample
completeness by $\sim10\%$.

We infer below that the performance achieved by the SVM$_{best}$
algorithm may represent the best possible classification of stars and
galaxies that could be done, based on single-epoch $ugriz$ photometry
alone.  However, it is unlikely that an ideal training set will be
available for object classification in future, deep datasets.
Identifying the regions of {\it ugirz} color--color space where
classification fails can highlight possible ways to improve the
unsupervised HB (or ML) classification methods implemented here.  For
example, we want to check for regions of color-color space in which
templates used in ML and HB may be missing, or to check whether the
implementation of simple, but stronger, priors could increase
performance.

\begin{figure}
\centering
 \includegraphics[clip=true, trim=0cm 0cm 0.0cm 0.cm,width=8cm]{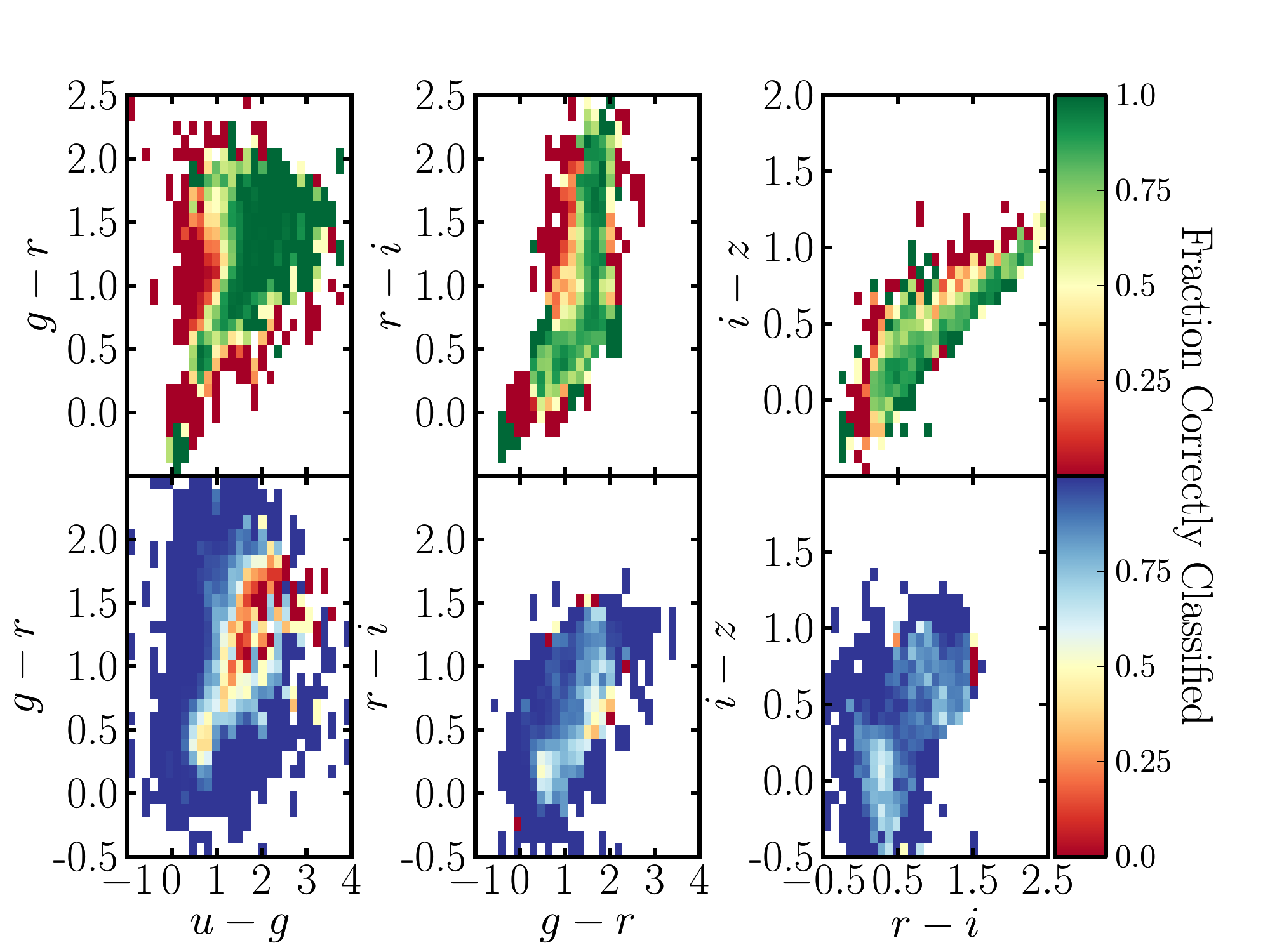}
\caption{The fraction of objects correctly classified at
  $\ln(\Omega)=0$ using our HB template fitting, distributed in {\it
    ugriz} color-color space.  The top panel shows the performance on stars, and the bottom panel shows the performance on galaxies.  Comparing with Figure
  \ref{fig:color-color-data}, it is clear that classification is most
  successful for regions in which stars and galaxies do not overlap in
  color-color space.}
\label{fig:color-color-hb-fraction}
\end{figure}

\begin{figure}
\centering
 \includegraphics[clip=true, trim=0cm 0cm 0.0cm 0.cm,width=8cm]{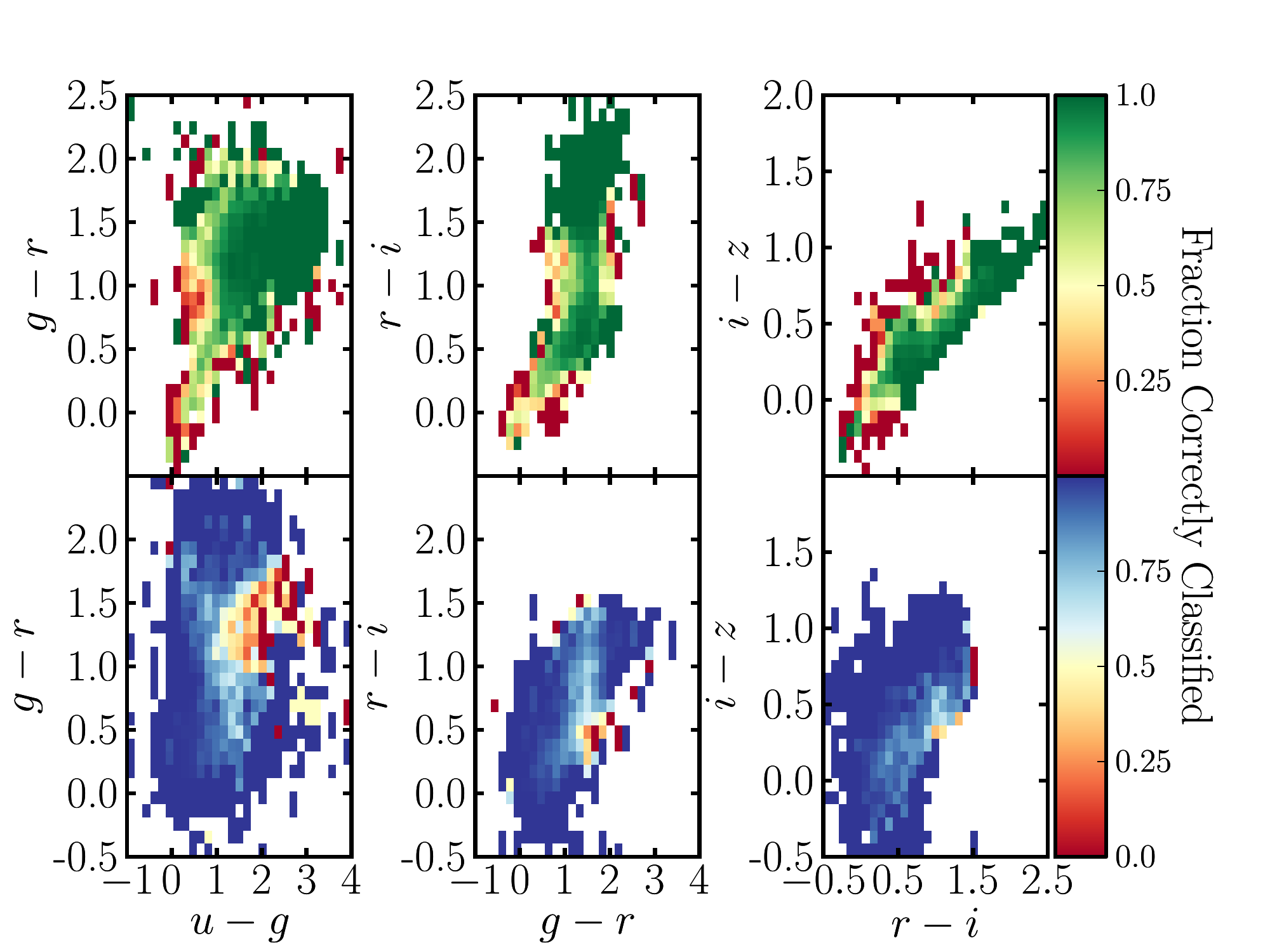}
\caption{The same as Figure \ref{fig:color-color-hb-fraction}, but for
  a SVM trained with data which span the $S/N$ range of the whole sample
  (SVM$_{best}$).  The top panel shows the performance on stars, and
  the bottom panel shows the performance on galaxies.  By inspection,
  it is clear that SVM$_{best}$ outperforms HB template fitting,
  particularly in the case of galaxies.  A striking difference is the
  poor galaxy classification of HB compared to SVM$_{best}$ in $u-g$.
  This may indicate a model deficiency in the $u$ spectral range of
  our galaxy templates.}
\label{fig:color-color-svm-fraction}
\end{figure}

Figures \ref{fig:color-color-hb-fraction} and
\ref{fig:color-color-svm-fraction} show the fraction of sources
correctly classified using HB and SVM$_{best}$, distributed over
colors.  Comparing with Figure \ref{fig:color-color-data} reveals that
the places where classification is least successful are regions where
stars and galaxies overlap the most in color.  For example, both the
SVM$_{best}$ and the HB algorithm struggle to correctly identify
galaxies with 1 $<$ $u-g$ $<$ 3 and 1 $<$ $g-r$ $<$1.5.  The number
density of galaxies in the failing region is low, making HB even more
likely to call everything a star.  Similarly, both stars and galaxies
populate $u-g < 1$ and $g-r \sim 1$, presenting a challenge to both
SVM and HB algorithms.  In this case, the number density of galaxies
is higher than that of stars, making HB even more likely to call
everything a galaxy and training SVM on a color separation that favors
galaxies over stars.

In the region of $r-i>1.5$, the stellar locus has essentially zero
overlap with galaxies in the sample.  The SVM$_{best}$ algorithm
yields exquisite classification of these stars, while the HB algorithm
returns only a mediocre performance (although $g-r < 1$ and $r-i >
1.5$ is populated with few stars, so those poorly classified regions
do not represent a significant fraction of all stars).  In future
work, the classification of $r-i > 1.5$ stars could therefore be
improved with the implementation of stronger priors on the permitted
redshifts at which galaxies may live---for example, by forcing a zero
probability of elliptical galaxies at high redshifts.

Locating regions of color space in which the classifiers struggle to
correctly separate stars and galaxies not only helps to decipher
weaknesses in classification algorithms, but can be used to identify
the specific science cases which will be most highly impacted.  To
illustrate, we examine places where both SVM$_{best}$ and HB
underperform and compare these regions to the object types in our
templates.  For stars we identify two such regions.  The first lies 
 within $0.0\lesssim u-g \lesssim 1.0$ and $0.7\lesssim g-r \lesssim
1.5$, which has been suggested to be comprised of white dwarf, M dwarf 
binaries \citep{silvestri06, covey07}.  The second region, with 
$0.0\lesssim u-g \lesssim 1.0$ and $0.0\lesssim g-r \lesssim 0.5$, is 
consistent with metal-poor main-sequence turnoff stars.  The relatively 
poorer performance in this region is particularly troubling, since these 
populations are some of the main tracers for low-surface brightness 
Galactic halo structure.  

For galaxies, association of underperforming regions to specific
populations is less clear-cut.  For instance, we find the poor
performing region with $1.5\lesssim u-g \lesssim 3.0$ consistent with
S0/SA SEDs with redshifts less than 0.4, but also with dusty
starbursting galaxies across a wider redshift range.  While far from
comprehensive, these associations highlight the fact that
classification performance can affect certain science cases more than
others, and should be accounted for both during individual analyses
and in future algorithm development.

\begin{figure}
\centering
 \includegraphics[clip=true, trim=0cm 0cm 0.0cm 0.cm,width=8cm]{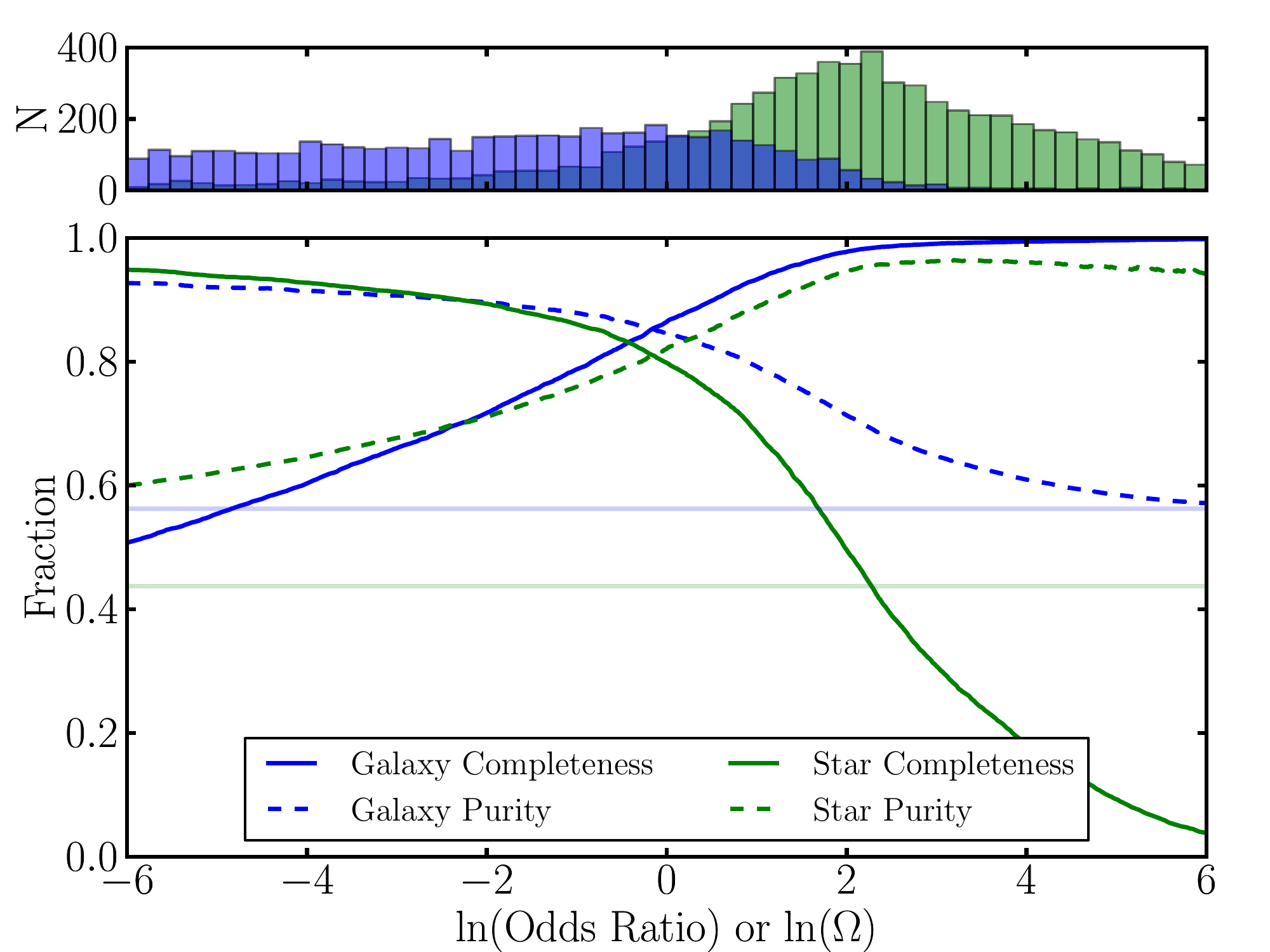}
\caption{Hierarchical Bayesian template fitting results showing
  completeness (solid) and purity (dashed) lines as a function of
  $\ln\Omega$.  Results for stars are shown in green and galaxies are
  shown in blue, while the solid (dashed) curves show completeness
  (purity).  Also indicated by green and blue horizontal lines is the
  relative fraction of stars and galaxies in the sample, respectively.
  The top panel shows the histograms associated with the
  distribution. Setting $\ln\Omega>=6$ effectively calls all sources
  galaxies, so the galaxy completeness is high, while the purity is
  set by the sample fraction of galaxies.  The same conclusions are
  reached for stars at $\ln\Omega<-6$.  The exact value of $\ln\Omega$
  chosen depends on the completeness and purity requirements dictated
  by the user's science case.}
\label{fig:hb-logodds}
\end{figure}

\begin{figure}
\centering
 \includegraphics[clip=true, trim=0cm 0cm 0.0cm 0.cm,width=8cm]{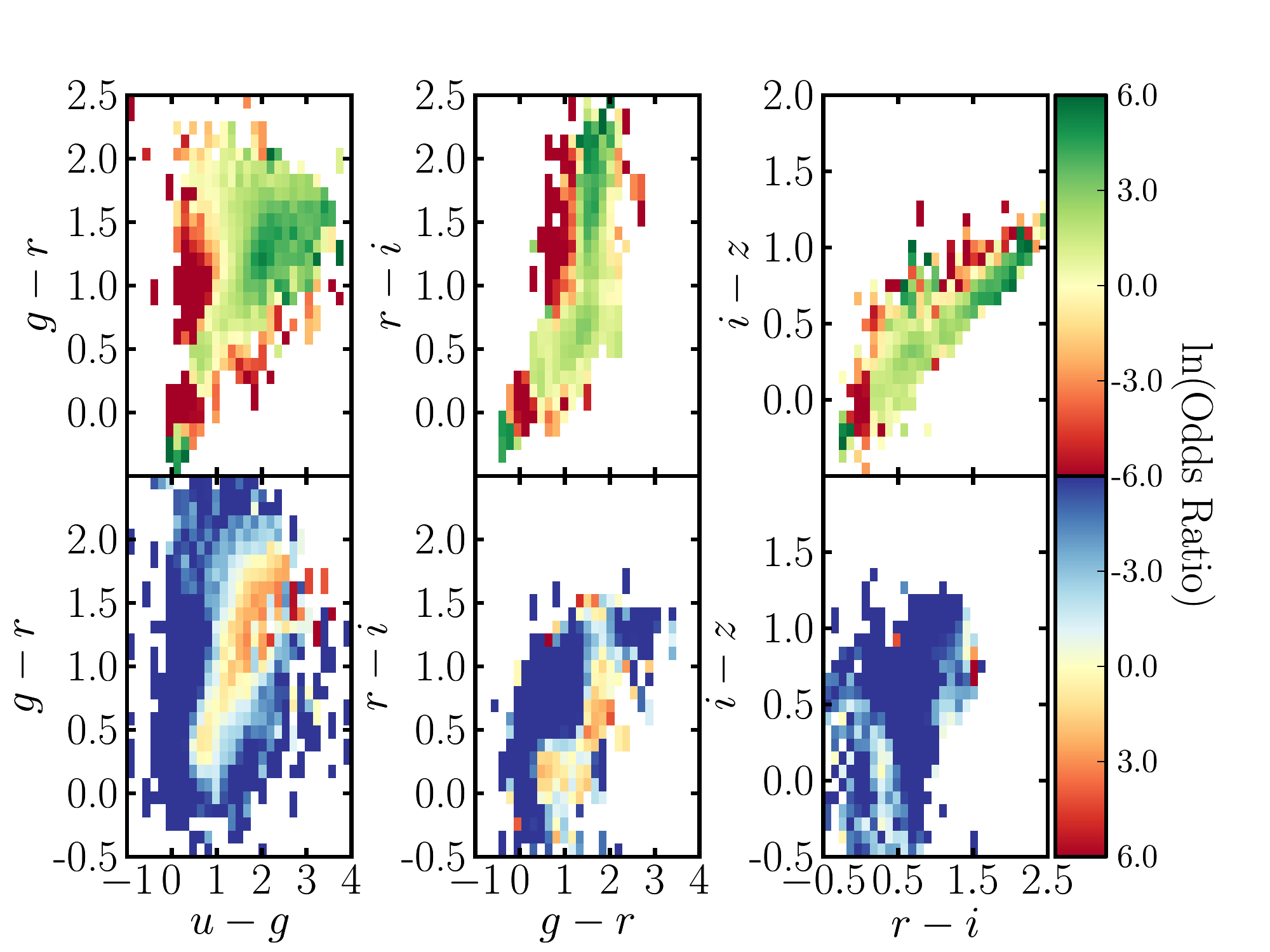}
\caption{The median $\ln(\Omega)$ of objects produced by our HB
  template fitting, distributed in {\it ugriz} color-color space.
  Similar to Figure \ref{fig:color-color-hb-fraction}, regions with
  the most extreme $\ln(\Omega)$ values are primarily those which have
  little color--color overlap between stars and galaxies.  While
  altering the $\ln(\Omega)$ threshold can deliver more pure or
  complete samples (cf. Figure \ref{fig:hb-logodds}), it may likely
  bias the sample to certain regions of color space.}
\label{fig:color-color-hb-odds}
\end{figure}

One of the great advantages of probabilistic classification is that
one need not restrict the classification criterion to a fixed value.
By moving away from $\ln(\Omega)=0$, one can obtain more/less pure or
complete samples of stars and galaxies, depending on the user's
science case.  In detail, how the purity or completeness varies as a
function of $\ln(\Omega)$ depends on the algorithm used.  To
illustrate, we show in Figure \ref{fig:hb-logodds} how purity and
completeness vary for the log odds ratio output by our HB algorithm.
In the figure, as $\ln(\Omega)$ decreases, we are requiring that the
relative likelihood that an object is a galaxy is much higher than
that for a star.  Similarly, as $\ln(\Omega)$ increases we are
requiring objects be more stringently classified as a star.  Thus, by
moving away from $\ln(\Omega)=0$ we change the star/galaxy purity and
completeness to the point where everything is called a star or galaxy,
giving 100\% complete samples with a purity set by the sample
fraction.  One caveat, however, is that modifying the threshold
$\Omega$ to achieve more pure samples may select objects which lie in
particular regions in SED space.  To illustrate, we show in Figure
\ref{fig:color-color-hb-odds} the distribution of $\ln(\Omega)$ in
color space.

We have considered the completeness and purity of sets of data
classified as stars or galaxies (as a function of $\ln(\Omega)$) as
one means of comparing different classification algorithms.  A
strength of this approach to quantifying the efficacy of
classification algorithms is its transparent connection to different
science requirements, in terms of purity and completeness.  A weakness
of this approach is the impossibility of selecting an overall ``best"
algorithm that presents an average over competing scientific
requirements.  For example, Figure \ref{fig:completeness} shows that
compared to SVM$_{best}$, our HB method gives better completeness in
stars but slightly worse completeness for galaxies---which performs
better in general?

\begin{figure}
\centering
 \includegraphics[clip=true, trim=0cm 0cm 0.0cm 0.cm,width=8cm]{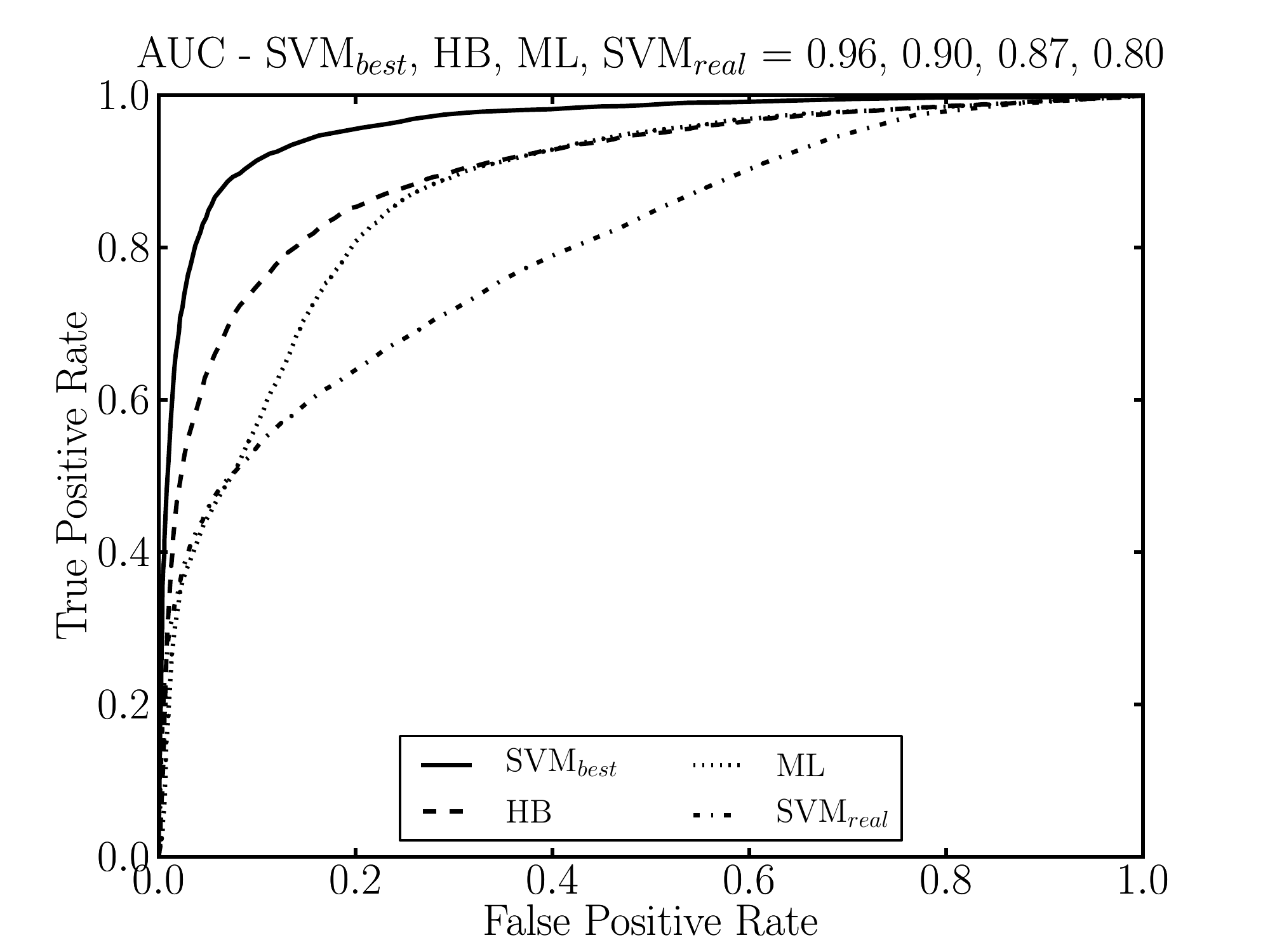}
\caption{The Receiver Operating Characteristic (ROC) curve for four photometric 
classification approaches: SVM$_{best}$, SVM$_{real}$, ML, and HB.  The ROC curve 
shows the true positive rate versus the false positive rate, as $\ln(\Omega)$ varies.  An ideal classifier 
always returns a true positive rate of one, so the Area Under the Curve (AUC) provides a 
general assessment of the performance.}
\label{fig:roc}
\end{figure}

We assess the overall performance of the various classification
algorithms using the Receiver Operating Characteristic (ROC) curve.  A
ROC curve is a plot of the true positive rate versus the false
positive rate of a binary classifier, as the classification threshold
($\ln(\Omega)$) is varied.  In Figure \ref{fig:roc}, we plot the ROC
curve for all four classification approaches considered here.  An
ideal classifier has a true positive rate equal to one for all values
of $\ln(\Omega)$.  Thus, the Area Under the Curve (AUC) statistic is
an assessment of the overall performance of the classifier.  There are
several points worth noting in Figure \ref{fig:roc}.  First, we find our HB
approach to template fitting outperforms the ML approach.
Considering our simple HB implementation is not very computationally
demanding (tens of minutes on typical desktop computer), even a basic
HB approach should always be preferred over the ML case.  SVM
algorithms, when trained with data which accurately capture the SED
and $S/N$ properties of the entire data, generally perform much better
than our current template fitting methods.  This is not surprising,
since template driven algorithms are never likely to have as complete
models as something data driven.  In reality, available training data will 
likely only capture the high
$S/N$ end of the survey in question.  As shown in Figure \ref{fig:roc},
a SVM$_{real}$ scenario underperforms even ML template fitting, 
casting severe doubt onto the usefulness of SVM with ill-suited
training information.  Future surveys which intend to use supervised 
techniques, therefore, will have to carefully consider if alternate strategies 
for obtaining training data \citep[e.g.,][]{richards12a,richards12b} can 
outperform template fitting methods.

%
%

\section{Conclusions}

Imminent and upcoming ground-based surveys are observing large
portions of the sky in optical filters to depths ($r\gtrsim24$),
requiring significant amounts of money, resources, and person power.
In order for such surveys to best achieve some of their science goals,
accurate star--galaxy classification is required.  At these new
depths, unresolved galaxy counts increasingly dominate the number of
point sources classified through morphological means.  To investigate
the usefulness of photometric classification methods for unresolved
sources, we examine the performance of photometric classifiers using
{\it ugriz} photometry of COSMOS sources with intrinsic FWHM
$<0.2$~arcsec, as measured with {\it HST}.  We have focused our
analysis on the classification of full survey datasets with broad
science goals, rather than on the classification of subsets of sources
tailored to specific scientific investigations.

Our conclusions are as follows:

\begin{itemize}

\item Maximum Likelihood (ML) template fitting methods are simple, and
  return informative classifications.  At $\ln(\Omega)=0$, ML methods
  deliver high galaxy completeness ($\gtrsim90\%$) but low stellar
  completeness ($\sim50\%$).  The purity of these samples range from
  $\sim50-95\%$, and are a strong function of the relative sample
  fraction.

\item We present a new, basic Hierarchical Bayesian (HB) approach to template
  fitting which outperforms ML techniques, as shown by the Receiver
  Operating Characteristic (ROC).  HB algorithms have no need for training, 
  and have nuisance parameters that are tuned according to the likelihood 
  of the data itself.  Further improvements to this basic algorithm are possible by
  hierarchically modeling the redshift distribution of galaxies, the
  SEDs of the input templates, and the distribution of apparent
  magnitudes.

\item Support Vector Machine (SVM) algorithms can deliver excellent
  classification, which outperforms template fitting methods.
  Successful SVM performance relies on having an adequate set of
  training data.  For optimistic cases, where the training data is
  essentially a random sample of the data (with known
  classifications), SVM will outperform template fitting.  In a
  more-realistic scenario, where the training data samples only the
  higher signal to noise sources in the data to be classified, SVM
  algorithms perform worse than the simplest template fitting methods.

\item It is unclear when, if ever, adequate training data
  will be available for SVM-like classification, HB algorithms are 
  likely the optimum choice for next-generation classifiers.

\item A downside of a paucity of sufficient training data is the
  inability to assess the performance of both supervised (SVM) and
  unsupervised (ML, HB) classifiers.  If knowing the completeness and
  purity in detail is critical to the survey science goals, it may be
  necessary to seek out expensive training/testing sets.  Otherwise,
  users will have to select the best unsupervised classifier (HB
  here), and rely on performance assessments extrapolated from other
  studies.

\item Ground based surveys should deliver probabilistic photometric 
 classifications as a basic data product.  ML likelihoods are useful and 
 require very little computational overhead, and should be considered the 
 minimal delivered quantities.  Basic or refined HB classifications require 
 more overhead, but can be run on small subsets of data to learn the 
 priors and then run quickly on the remaining data, making them a feasible 
 option for large surveys.  Finally, if excellent training 
 data is available, SVM likelihoods should either be computed or the data 
 should be made available.  In any scenario, we strongly recommend that 
 likelihood values, not binary classifications, should be delivered so that 
 they may be propagated into individual analyses.

\end{itemize}

The future of astronomical studies of unresolved sources in ground
based surveys is bright.  Surveys like PanSTARRS, DES, and LSST will
deliver data that, in conjunction with approaches discussed here, will
expand our knowledge of stellar systems, the structure of the Milky
Way, and the demographics of distant galaxies.  We have identified
troublesome spots for classification in single-epoch $ugriz$
photometric data, which may hinder studies of M-giant and metal-poor
main-sequence turnoff stars in the Milky Way's halo.  Future studies
could improve upon our preliminary results by impleneting
more-sophisticated prior distributions, by identifying crucial
improvements needed in current template models or training data, or by
pursuing complementary non-SED based classification metrics.

\acknowledgments

We gratefully acknowledge P. Capak and the COSMOS team for providing
an up-to-date version of their catalog, C.-C. Chang and C.-C. Lin for
making their SVM routines available, J. J. Bochanski for providing his 
stellar templates, and the {\tt Le Phare} photo-z
team for making code and templates available.  We wish to thank J.
Newman, P. Thorman, S. J. Schmidt, D. Foreman-Mackey, and M. Juric 
for helpful and insightful conversations.  A special thanks is owed to 
\u{Z}. Ivezi\'{c} and P. Yoachim for leading us to an improved understanding 
of the COSMOS classifications.  We also thank Joe Cammisa, Mulin 
Ding, and Dustin Lang for technical support.  RF and BW also thank 
the NYU Center for Cosmology and Particle Physics, and 
Drexel University's Physics Dept. for hosting them during the writing of 
this paper.  RF and BW acknowledge support from NSF grant 
AST-0908193.  DH acknowledges support from the NSF grant IIS-1124794.

\clearpage

\bibliographystyle{apj}

\appendix
\section{Hierarchical Bayesian Star--Galaxy Classification}

In this appendix, we provide a detailed, step-by-step description of our Hierarchical 
Bayesian algorithm.  First, let us define the data as the sets:
\footnotesize
\begin{eqnarray}\displaystyle
\flux & = & \{10^{-\frac{2}{5}m_1}F_{1,0},\,...\,,10^{-\frac{2}{5}m_l}F_{l,0},\,...\,,10^{-\frac{2}{5}m_N}F_{N,0}\}
\nonumber\\
\uncertainty & = & \{\frac{2}{5}\ln(10)F_1\sigma_{m_1},\,...\,,\frac{2}{5}\ln(10)F_l\sigma_{m_l},\,...\,,\frac{2}{5}\ln(10)F_N\sigma_{m_N}\}
\quad ,
\end{eqnarray}
\normalsize
\noindent where $m_l$, $\sigma_{m_l}$ is the observed magnitude and 
uncertainty in filter number $l$ for $N$ number of filters.  One  
sequence for the filters $l$ corresponds to $\{l\}=\{u,g,r,i,z\}$. The 
zeropoint, $F_{l,0}$, is:
\begin{eqnarray}\displaystyle
F_{l,0}=\int \lambda \,S_\lambda\,R_{\lambda, l} \dd \lambda
\quad ,
\end{eqnarray}
where $S_\lambda$ is the standard flux density spectrum (Vega or AB), and 
$R_{\lambda,i}$ is the fraction of photons incident on the top 
of the atmosphere which are counted by the detector, as a function 
of wavelength.

Next, we generate a model for the data using the templates:
\begin{eqnarray}\displaystyle
F_{{\rm mod},l} & = & \int  \lambda\,f_{\lambda,{\rm mod}}\,R_{\lambda,l} \dd \lambda\,
\quad ,
\end{eqnarray}
where $f_{\lambda,{\rm mod}}$ corresponds to the flux density 
of a given spectral template.  Finally, we define a goodness of 
fit statistic:
\begin{eqnarray}\displaystyle\label{eqn:chi}
\chi^2 & = & \sum\limits_{l=1}^N \frac{(F_l-C_{\rm mod}\,F_{{\rm mod},l})^2}{\sigma_{{\rm total}_l}^2}
\quad ,
\end{eqnarray}
where $C_{\rm mod}$ is a constant unitless amplitude applied to the 
model for the fit (discussed more below as $C_{ij}$).  The variance 
$\sigma_{{\rm total}_l}^2 = \sigma_{F_l}^2 + \eta^2 F_{l}^2$, where $\eta$ is 
a few percent and represents a nuisance parameter which (in a global sense) accounts for 
error in the models as well as underestimates in $ \sigma_{F_l}^2 $.  The 
value of $\chi^2$ from our template fitting is the fundamental quantity 
on which our inference procedure is based, as follows below.

We represent the hypothesis that an object $i$ is a star or a galaxy
by ``$S$'' or ``$G$'' respectively.  For a given object $i$, we fit a
set of templates $j$ corresponding to $S$ using the procedure outlined
above.  The likelihood of template $j$ and amplitude $C_{ij}$ under the stellar
hypothesis $S$ given the single observed data point $\flux_i$ is:
\begin{eqnarray}\displaystyle
p(\flux_i|C_{ij},j,S) & \propto & \exp(-\frac{1}{2}\,\chi^2)
\quad ,
\end{eqnarray}
where $\flux_i$ is the full set of observations of object $i$ and the
associated noise model, and $\chi^2$ is defined in Equation \ref{eqn:chi}. Note
that the $\chi^2$ is not necessarily the best-fit value for $\chi^2$
but rather the $\chi^2$ obtained with template $j$ when it is given
amplitude $C_{ij}$.

We could optimize this likelihood, but really we want to compare the
whole $S$ model space to the whole $G$ model space.  We must
marginalize this likelihood over the amplitude and template.  To 
demonstrate this, let us step through each marginalization for the 
$S$ model space.

Marginalization over the amplitude $C_{ij}$ looks like
\begin{eqnarray}\displaystyle
p(\flux_i|j,S,\hyperpars) & = & \int p(\flux_i|C_{ij},j,S)\,p(C_{ij}|j,S,\hyperpars)\,\dd C_{ij}
\quad ,
\label{eqn:fitmarg}
\end{eqnarray}
where the integral is over all permitted values for the amplitude
$C_{ij}$, and the prior PDF $p(C_{ij}|j,S,\hyperpars)$ depends on the
template $j$, the full hypothesis $S$.  Note, the prior PDF obeys the normalization
constraint
\begin{eqnarray}\displaystyle
1 & = & \int p(C_{ij}|j,S,\hyperpars)\,\dd C_{ij}
\quad .
\label{eqn:fitconstraint}
\end{eqnarray}
\noindent Here we have also introduced some 
``hyperparameters'' $\hyperpars$, which are variables which parameterize 
prior distributions.  The subset of hyperparameters $\hyperpars$ which 
apply to $p(C_{ij}|j,S,\hyperpars)$ might be, for example, the mean and variance of 
a log-normal distribution on $C_{ij}$.   It is the simultaneous inference
of the star--galaxy probabilities and the hyperparameters that
make the approach hierarchical.  

Any realistic prior PDF for the $C_{ij}$ comes from noting that (for
stars), the $C_{ij}$ are dimensionless squared distance ratios between
the observed star and the template star; in this case the prior
involves parameters of the stellar distribution in the Galaxy.  When
we look at galaxies (below), this situation will be different.  In the
(rare) case that the prior PDF $p(C_{ij}|j,S,\hyperpars)$ varies
slowly around the best-fit amplitude,
\begin{eqnarray}\displaystyle
p(\flux_i|j,S,\hyperpars) & \propto & \exp(-\frac{1}{2}\,\tilde{\chi}^2)
  \,p(\tilde{C}_{ij}|j,S,\hyperpars)\,\sigma_{Cij}
\quad ,
\end{eqnarray}
where $\tilde{\chi}^2$ is the best-fit chi-squared, $\tilde{C}_{ij}$
is the best-fit amplitude, and $\sigma_{Cij}$ is the standard
uncertainty in $\tilde{C}_{ij}$ found by least-square fitting.  This
approximation is that the prior doesn't vary significantly
within a neighborhood $\sigma_{Cij}$ of the best-fit amplitude.  

Marginalization over the template space looks like
\begin{eqnarray}\displaystyle
p(\flux_i|S,\hyperpars) & = & \sum_j p(\flux_i|j,S)\,P(j|S,\hyperpars)
\quad ,
\label{eqn:tempmarg}
\end{eqnarray}
where $P(j|S,\hyperpars)$ is the prior probability (a discrete
probability, not a PDF) of template $j$ given the hypothesis $S$ and
the hyperparameters $\hyperpars$.  It obeys the normalization
constraint
\begin{eqnarray}\displaystyle
1 & = & \sum_j P(j|S,\hyperpars)
\quad .
\label{eqn:tempconstraint}
\end{eqnarray}
\noindent Note $P(j|S,\hyperpars)$ is a discrete set of weights, 
whose value corresponds to the hyperparameter for template $j$.

To summarize, the marginalized likelihood $p(\flux_i|S,\hyperpars)$ that 
a source $i$ is a star $S$ is computed as:

\begin{eqnarray}\displaystyle
\label{eqn:starmarg} 
p(\flux_i|C_{ij},j,S) & \propto & \exp(-\frac{1}{2}\,\chi^2)
\nonumber\\
p(\flux_i|j,S,\hyperpars) & = & \int p(\flux_i|C_{ij},j,S)\,p(C_{ij}|j,S,\hyperpars)\,\dd C_{ij}
\nonumber\\
p(\flux_i|S,\hyperpars) & = & \sum_j p(\flux_i|j,S,\hyperpars)\,P(j|S,\hyperpars)
\quad .
\end{eqnarray}

The marginalized likelihood that source $i$ is a galaxy $G$, is calculated 
following a very similar sequence.  In calculating the likelihood, we 
allow a given galaxy template $k$ to be shifted in wavelength by 
a factor $1+z$.  This introduces another step in the calculation that 
marginalizes the likelihood across redshift for a template, giving 

\begin{eqnarray}\displaystyle
\label{eqn:galmarg}
p(\flux_i|C_{ikz},k,z,G) & \propto & \exp(-\frac{1}{2}\,\chi^2)
\nonumber\\
p(\flux_i|k,z,G,\hyperpars) & = & \int p(\flux_i|C_{ikz},k,z,G)\,p(C_{ikz}|k,z,G,
					\hyperpars)\,\dd C_{ikz}
\nonumber\\
p(\flux_i|G,k,\hyperpars) & = & \sum_z p(\flux_i|k,z,G)\,P(z|k,G,\hyperpars)
\nonumber\\
p(\flux_i|G,\hyperpars) & = & \sum_k p(\flux_i|k,G)\,P(k|G,\hyperpars)
\quad ,
\end{eqnarray}
where now $C_{ikz}$ is the constant amplitude for galaxy template $k$
at a redshift $z$.  The marginalization across redshift also introduces 
a prior $P(z|k,G,\hyperpars)$, which is also is parameterized by a subset 
of $\hyperpars$, under some assumed form for the prior.

This model is fully generative; it specifies for any observed flux
$\flux_i$ the PDF for that observation given either the star
hypothesis $S$ or the galaxy hypothesis $G$.  We can write down then
the full probability for the entire data set of all objects $i$:
\begin{eqnarray}\displaystyle
\label{eqn:fullprob}
p(\allflux|\hyperpars) = \prod_i \left[p(\flux_i|S,\hyperpars)\,p(S|\hyperpars)
                                     + p(\flux_i|G,\hyperpars)\,p(G|\hyperpars)
                                \right]
\quad ,
\end{eqnarray}
where even the overall prior probability $p(S|\hyperpars)$ that an
object is a star (or, conversely, a galaxy) depends on the
hyperparameters $\hyperpars$.  These obey the normalization constraint
\begin{eqnarray}\displaystyle
1 = p(S|\hyperpars) + p(G|\hyperpars)
\quad .
\label{eqn:SGpriorconstraint}
\end{eqnarray}
The likelihood $p(\allflux|\hyperpars)$ is the total, marginalized
likelihood for the combined data set of all the observations $\flux_i$
for all objects $i$.  From here we can take a number of approaches. 
One option is to find the hyperparameters that maximize this total 
marginalized likelihood, or we can assign a prior PDF $p(\hyperpars)$ 
on the hyperparameters, and sample the posterior PDF in the 
hyperparameter space.  For computational reasons, we choose to 
optimize $p(\allflux|\hyperpars)$ in this work.

With either a maximum-likelihood set of hyperparameters $\hyperpars$
or else a sampling, inferences can be made.  For our purposes, the
most interesting inference is, for each object $i$, the posterior
probability ratio (or odds) $\Omega_i$
\begin{eqnarray}\displaystyle
\Omega_i & \equiv & \frac{p(S|\flux_i,\hyperpars)}
                         {p(G|\flux_i,\hyperpars)}
\nonumber\\
p(S|\flux_i,\hyperpars) & = & p(\flux_i|S,\hyperpars)\,p(S|\hyperpars)
\nonumber\\
p(G|\flux_i,\hyperpars) & = & p(\flux_i|G,\hyperpars)\,p(G|\hyperpars)
\quad ,
\end{eqnarray}
where we have re-used most of the likelihood machinery generated
(above) for the purposes of inferring the hyperparameters.  That is,
the star--galaxy inference and the hyperparameter inferences proceed
simultaneously.

\end{document}